# Artificial Intelligence Nomenclature Identified From Delphi Study on Key Issues Related to Trust and Barriers to Adoption for Autonomous Systems

Thomas E. Doyle*[†‡], *Senior Member, IEEE*, Victoria Tucci[§], Calvin Zhu[†], *Student Member, IEEE*, Yifei Zhang*, Basem Yassa*, Sajjad Rashidiani*, Md Asif Khan*, Reza Samavi[τ‡], *Senior Member, IEEE,*
Michael Noseworthy*[†κ], Steven Yule[σ]

*Abstract—* **The rapid integration of artificial intelligence across traditional research domains has generated an amalgamation of nomenclature. As cross-discipline teams work together on complex machine learning challenges, finding a consensus of basic definitions in the literature is a more fundamental problem. As a step in the Delphi process to define issues with trust and barriers to the adoption of autonomous systems, our study first collected and ranked the top concerns from a panel of international experts from the fields of engineering, computer science, medicine, aerospace, and defence, with experience working with artificial intelligence. This document presents a summary of the literature definitions for nomenclature derived from expert feedback.**

*Index Terms—* artificial intelligence, cross-discipline, definitions machine learning, nomenclature.

## I. Introduction

THE incorporation of artificial intelligence (AI) into areas of decision support, team support, and human-AI partnership requires diverse perspectives from multiple domains of research, development, and operationalization. To improve the understanding of issues of trust and barriers to the adoption of such systems, an expert Delphi panel was consulted. The initial stage of the Delphi process was to gather expert feedback on the top three challenges regarding the adoption and implementation of AI and the top three key issues related to trust in AI. After receiving feedback from the initial stage, the 248 concepts were organized into 33 groups, with each group represented by a specific term or phrase derived directly from the responses, thus generating a nomenclature. This nomenclature list will be used for the next stages of the Delphi to gain consensus on the importance and feasibility of each term.

The following are the top-ranked terms that have definitions derived from the literature (listed alphabetically):

- AI Integration Process and Standards
- Accuracy of Results/Output
- Data Veracity
- Ethical Implementation and Fairness
- Feedback to User and Communications to/from AI
- Flexibility and Adaptability to Novel Situations and Unexpected Stimuli
- Government Regulations and AI Policy
- Model Bias
- Operator - AI Interaction and Teaming
- Performance
- Predictability of AI Behaviour
- Privacy
- Reliability
- Robustness
- Security
- System and/or Model Errors and Failures
- Testing and Validation
- Transparency
- Trust
- Unbalanced Dataset and Dataset Shift
- Understanding AI Capabilities / AI Education
- Usefulness/Benefit to the User

The list of terms will be used as a common reference for the

*This work was funded and supported by the Canadian Department of National Defence IDEaS under award number CFPMN2-17 and the Department of Electrical and Computer Engineering at McMaster University, Canada. Corresponding author: T. E. Doyle mail:doylet@mcmaster.ca*

Author affiliations: *Department of Electrical and Computer Engineering, McMaster University, Hamilton, Canada

[†]School of Biomedical Engineering, McMaster University, Hamilton, Canada

[κ]Department of Imaging Physics and Engineering, Imaging Research Centre, St. Joseph's Healthcare, Hamilton, Ontario, Canada

[‡]Vector Institute for Artificial Intelligence, Toronto, Canada

[§]Faculty of Health Sciences, McMaster University, Hamilton, Canada

[τ]Department of Electrical, Computer, and Biomedical Engineering, Toronto Metropolitan University, Toronto, Ontario, Canada

[σ]Department of General Surgery, Royal Infirmary, University of Edinburgh, Edinburgh, EH16 4SA, UK

next stages of the Delphi study. It should be noted that for many of the definitions, the literature is not consistent and can be ambiguous. The authors have provided succinct definitions based on their experience with the subject and present them here as the basis for further research and discussion.

II. NOMENCLATURE

This section contains the Delphi-derived nomenclature definitions. Each definition is supported by several citations from the literature, and where appropriate, overlaps with other definitions are noted. The order presented here matches the randomized list order provided to the Delphi participants. An alphabetical list was presented in the Introduction.

*AI Integration Process and Standards*

Artificial Intelligence (AI) integration and standards refer to the structured integration of intelligent algorithms into the specific domain workflow. From data acquisition, to labelling, to modelling, to end-user interaction, research continues on the operationalization of AI [1]. While methods have been proposed, integration processes and standards are still evolving and in development; however, there do exist standards for low-level automation functions [2] [3] [4] [5]. For reference, a global repository of current and emerging AI standards demonstrates the evolving work in the field [4]. IEEE 2801-2022 is the only recommended practice we are aware of which addresses the quality management of datasets for the medical artificial intelligence [6]. ISO is also in the process of developing ISO/IEC JTC 1/SC 42 for AI [7]. According to ISO, standardization in this area provides guidance to ISO committees developing Artificial Intelligence applications.

*Testing and Validation*

Testing and validation are often used as a reference to development stages where trained models are evaluated using with-held data. This specific development stage is also referred to as 'internal evaluations'. It is also important to consider external validation which, instead, refers to external evaluations to be made on the models beyond the with-held data. Formally, by testing, we mean evaluating the system in several conditions and observing its behavior, watching for defects. Likewise, by verification, we mean producing a compelling argument that the system will not misbehave under a very broad range of circumstances [8] [9].

*Reliability*

Reliability of AI is the demonstrated capacity of AI, based on its previous performance, in accomplishing specific tasks [10]. In other words, AI reliability in a task is measured by analyzing performance data from previous instances where AI has been utilized in the same task [10].

*Trust*

Trust in AI agents is defined in the context of delegation without supervision. It is the delegators' attitude towards AI when assigning an AI agent to a place that has always been occupied by human trustees [11]. Trust is the delegators' positive expectation from the AI agent's future performance [10].

*System and/or Model Errors and Failures*

In machine learning (ML), model errors are one of the main impediments to improved accuracy and reliability as errors can propagate, become amplified, or be suppressed, making it challenging to find the reason(s) for failure [12] [13]. ML model error generally denotes incorrect prediction for classification tasks, whereas, in the case of regression tasks, it is considered a significant deviation from the ground truth that results in a failed ML system [14]. The consequences of these failures reduce model reliability, safety, and trust in ML [15]. Therefore, the common sources of errors need to be identified to reduce or prevent failures. Two major model-associated errors are faulty model assumptions and model fragility. For example, a linear ML model may be inappropriate and produce errors where nonlinearities are among the features [12].

*Model Bias*

There are several sources of bias, starting from data collection to model deployment, that can affect the performance and implications of an ML system [13] [16]. The bias observed in a resulting or final ML/AI model is model bias or sometimes termed algorithmic bias [17] [18] [19] [20]. Bias in the outcome of the model is caused by bias propagation via the ML pipeline. For example, bias in the data collection stage may influence an object classification model to learn features of the object only in certain conditions, or underrepresented groups in medical research could be negatively affected due to the model bias [16] [18].

*Feedback to User and Communications to/from AI*

AI feedback/communication to the user can be defined as the results produced by the AI model which conclusions can be drawn from. These results can include, but are not limited to, predictions, statistics, and even raw values. User feedback/communication to AI is the process of adapting user input to retrain or adjust features in an ML model. The back-and-forth communication creates a feedback-control loop which is part of the process known as reinforcement learning [21] [22] [23] [24].

*Accuracy of Results/Output*

In ML, accuracy is a term used mainly for evaluating supervised ML algorithms. The predictive power of a classification algorithm, or how well a classifier can predict the hold-out test data, is typically measured by its predictive accuracy or error rate (1 - accuracy) [25] [26] [27]. It is the ratio of the number of correct predictions to the number of total predictions. It is a useful measure when all classes considered in a classification problem are of equal importance [28].

*Ethical Implementation and Fairness*

There has been an increasing push for "ethical AI", but it currently has varying definitions. Some consider ethical implementation to be properties, with the most common being transparency, justice and fairness, non-maleficence, responsibility, and privacy. Others refer to ethical considerations taken at the various steps of development to prevent unintended bias and discrimination. For example, there may be inherent biases in the training dataset. These biases are reflected in model outputs and may lead to unanticipated issues, such as racial discrimination. Other examples are financial incentives and privacy concerns regarding data collection and use [29] [30] [31].

An algorithm is said to be unfair if a pair of inputs that are close together are treated differently. An algorithm is said to be fair if no such pair exists. "Close" is determined by an appropriately defined metric and differing outputs may be vastly differing class labels or other such large differences in output [32].

*Operator - AI Interaction and Teaming*

Operator – AI interaction is the process that aims to give the user/operator the ability to control and coordinate artificial intelligence. This can be done through a collection of technologies, such as wearable devices and sensors, as well as the fusion of machine intention recognition with the user emotion recognition. This helps provide the user/operator with a collaborative environment where they can be an active participant in the process [33] [34].

*Unbalanced Data and Dataset Shift*

Other common names for unbalanced data:
- Imbalanced data
- Class sparsity
- Class imbalance

Most ML algorithms assume uniform class size, however, in reality this is often not the case. Imbalanced datasets are datasets in which there is an unequal distribution between its classes. By this definition, the majority of datasets are imbalanced, however it is more common to use this term for datasets that have classes that differ greatly in size. This can lead to skewed accuracies where models tend to predict the majority class [35] [36] [37].

Other names for dataset shift:
- Concept shift, concept drift
- Covariate shift
- Changing environments

Dataset shift is common as the underlying distribution of the modelled data is often unknown during the training process and even if it is known, the distribution may change between training and deployment. Suppose there is a classification problem defined as predicting outcome Y from feature X which comes from data collected from distribution A. Suppose that during deployment, the data will be collected from distribution B. We describe this the differences between the distributions of the training data and deployment data as dataset shift. The differing distributions can be due to several factors, such as demographics or sensor resolution [38].

*Predictability of AI Behavior*

Unpredictability, also known as unknowability or cognitive uncontainability, is the ability to precisely and consistently predict AI's action to achieve the objective [39]. Predictability relates to the question of "what will the system do?"; in an AI behavior sense, it is more of the accuracy of AI's execution of a task, in relation to its training data, and how well it can accommodate to input data-- in other words, if the behavior can be anticipated. It is different from reliability (the extent to which the system does or does not fail), as failing could be predictable. Predictability could be related to training data, testing against variety of inputs, complexity and quality of input data, self-learning ability, etc. [40].

*Performance*

Performance of different AI algorithms can be defined as how well they complete their tasks and the confidence levels for predictions in some situations, which means for every AI the performance metrics should be varied. Some of the metrics that exist are regression metrics, classification accuracy, confusion matrix of ground-truth labels versus model predictions, precision, recall, F1-score, etc. [41] [42] [43].

*Robustness*

Robustness in ML can refer to multiple facets of an ML pipeline. Generally, robustness in ML refers to model robustness, which denotes the degree of effectiveness of a model in generalizing beyond its training data ranging from the natural shift in data distribution to tackling adversaries. It is also referred to as model stability, and for a model to be robust, the performance deviation should be minimal [44] [45] [46] [47]. Models' diversity and heterogeneity are considered the key to building robust and accurate models [48]. Additionally, a robust model should behave as expected under likely, unlikely, anomalous, or adversarial conditions; with re-producible empirical behavior [46]. Model robustness is also related to data quality, model decay, feature stability, precision vs. recall, and input perturbations [45]. For example, a robust neural network image classifier should not get fooled while testing it by perturbing a test image [49]. This type of attack to ML is usually called adversarial examples. Robustness can also be impacted by poisoning attacks where the decision boundary is gradually shifted in an unnoticeable way by an adversary in training time [50].

*Usefulness/Benefit to the User*

Usefulness of AI to the user is the evaluation of the advantages and positive impacts that AI provides in different services that traditionally require human support, such as checking symptoms based on patient data [51] [52].

*Government Regulations and AI Policy*

Government regulations and AI policy is a broad topic including ethics codes, principles, guidelines, frameworks, and policy strategies. Specifically in Canada, the Directive on Automated Decision-Making oversees the assistance of AI in making administrative decisions to improve service delivery. Other regulations focus on funding to attract and retain AI researchers. Recently on June 16, 2022, House of Commons passed a bill C-27 to regulate international and interprovincial trade and commerce in AI, by requiring measures to mitigate risks of harm and biased output related to high-impact AI systems [53] [54] [55].

*Transparency*

The exact usage of 'transparency' is unclear. It is sometimes used interchangeably or in tandem with 'explainability' and/or 'interpretability'.

Machine learning, in particular deep learning, is often described as a 'black box' where decisions are not easily interpreted or explained. Recent trends are pushing for more 'transparent' AI with IEEE P7001 being proposed as a standard for transparency in autonomous systems. There are varying definitions, but IEEE P7001 defines transparency as "the transfer of information from an autonomous system or its designers to a stakeholder, which is honest, contains information relevant to the causes of some action, decision or behavior and is presented at a level of abstraction and in a form meaningful to the stakeholder" [56] [57] [58].

*Security*

AI security may be defined as preventing, identifying, and neutralizing various types of adversarial attacks that can target AI systems in various phases of their development and cause deficiencies in their performance [59].

*Privacy*

Privacy is defined as one's right to have full control and secrecy over their personal information [60]. Violation of privacy in an AI system happens when a wrong party gets access or reliably infers individuals' sensitive information, the process of accessing information is compromised, or the access is obtained for wrong purposes [61]. Privacy-preserving AI offers solutions for countering these violations by implementing various techniques ranging from Federated Learning, Differential Privacy, Homomorphic Encryption and Secure Multi-Party Computation [10] [11] [59] [62] [63].

*Understanding AI Capabilities / AI Education*

In reference to the end-user, this is understood as general improvement to literacy of artificial intelligence concepts and capabilities. By understanding, this could improve trust of the technology.

*Data Veracity*

Data veracity is often equated with data trustworthiness because it is so closely bound to inherent uncertainty [64] [65]. As the desire for deep learning approaches grows, so too does the need for larger datasets. With the increasing size of datasets there can be biases, ambiguities, variances, errors, and gaps that need to be identified and addressed to improve the accuracy and generalizability of the models [65]. In the context of artificial intelligence, data veracity is the quality or integrity of the data to be used for training models [66]. IEEE 2801-2022 is the only recommended practice we are aware of which addresses quality management of datasets for medical artificial intelligence [6].

*Flexibility and Adaptability to Novel Situations and Unexpected Stimuli*

Novel or unexpected stimuli are cases where input to the trained artificial intelligence lies outside the scope of data of which the artificial intelligence was trained. Using the analogy of a 3-variable (A, B, C) Venn diagram, where our AI model was trained under the union A + B + C, the unexpected falls in A'B'C' (everything that is not A + B + C). Flexibility or adaptability are not explicit attributes in current modelling development methods. The current approach to such instances is to make a decision model more generalizable, thus broadening the scope of training by including the novel data in re-training. Beyond generalizability, out-of-distribution (OOD), or dataset shift, detection is an indicator that the modelling is receiving unexpected stimuli [67] [68]. In current literature the concepts of flexibility and adaptability are regularly referred to as agility, adaptiveness, robustness, and resilience; however, these are generally detection of anomalies and not truly a model's ability to adapt real-time [69].

V. DISCUSSION

This paper presents an expert panel's view on the most important concepts surrounding trust and barriers to adoption of autonomous systems. The terms presented form a common reference for ongoing discussion and research in the next phase of the Delphi process.

ACKNOWLEDGMENT

The authors would like to thank the expert participants of the Delphi for their valuable insights and feedback, the broader McMaster Micro-Net for their guidance.

REFERENCES

[1] Y. Roh, G. Heo, and S. E. Whang, "A Survey on Data Collection for Machine Learning: A Big Data - AI Integration Perspective," *IEEE Transactions on Knowledge and Data Engineering*, vol. 33, no. 4, pp. 1328–1347, Apr. 2021, doi: 10.1109/TKDE.2019.2946162.

[2] A. Sutton, R. Samavi, T. E. Doyle, and D. Koff, "Method for enabling trust in collaborative research," US11244076B2, Feb. 08, 2022 Accessed: Oct. 12, 2022. [Online]. Available: https://patents.google.com/patent/US11244076B2/en

[3] "ISO - ISO/IEC JTC 1/SC 42 - Artificial intelligence." https://www.iso.org/committee/6794475/x/catalogue/ (accessed Oct. 12, 2022).

[4] "The Global AI Standards Repository," *OCEANIS*. https://ethicsstandards.org/repository/ (accessed Oct. 12, 2022).

[5] "IEEE Recommended Practice for Industrial Agents: Integration of Software Agents and Low-Level Automation Functions," *IEEE Std 2660.1-2020*, pp. 1–43, Jan. 2021, doi: 10.1109/IEEESTD.2021.9340089.

[6] IEEE, "IEEE Recommended Practice for the Quality Management of Datasets for Medical Artificial Intelligence".

[7] "ISO/IEC JTC 1/SC 42 - Artificial intelligence," *iTeh Standards Store*, 2022. https://standards.iteh.ai/catalog/tc/iso/a8b53a70-2bb4-40a8-abf1-f42dde4432c5/iso-iec-jtc-1-sc-42 (accessed Oct. 14, 2022).

[8] F. Cabitza *et al.*, "The importance of being external. methodological insights for the external validation of machine learning models in medicine," *Comput Methods Programs Biomed*, vol. 208, p. 106288, Sep. 2021, doi: 10.1016/j.cmpb.2021.106288.

[9] Ian Goodfellow and Nicolas Papernot, "The challenge of verification and testing of machine learning," *cleverhans-blog*, Jun. 14, 2017. http://cleverhans.io/security/privacy/ml/2017/06/14/verification.html (accessed Oct. 06, 2022).

[10] Y. Hong *et al.*, "Statistical perspectives on reliability of artificial intelligence systems," *Quality Engineering*, vol. 0, no. 0, pp. 1–23, Jun. 2022, doi: 10.1080/08982112.2022.2089854.

[11] F. Fossa, "'I don't trust you, you faker!' On trust, reliance, and artificial agency," *Teoria*, vol. XXXIX, pp. 63–80, Jan. 2019, doi: 10.4454/teoria.v39i1.57.

[12] S. Saria and A. Subbaswamy, "Tutorial: Safe and Reliable Machine Learning." arXiv, Apr. 15, 2019. Accessed: Oct. 06, 2022. [Online]. Available: http://arxiv.org/abs/1904.07204

[13] N. Mehrabi, F. Morstatter, N. Saxena, K. Lerman, and A. Galstyan, "A Survey on Bias and Fairness in Machine Learning," *ACM Comput. Surv.*, vol. 54, no. 6, p. 115:1-115:35, Jul. 2021, doi: 10.1145/3457607.

[14] "Model error analysis," *Dataiku*, 2022. https://doc.dataiku.com/dss/latest/machine-learning/supervised/model-error-analysis.html (accessed Oct. 06, 2022).

[15] M. Sameki, L. Gayhardt, and Alex Buck, "Assess errors in machine learning models - Azure Machine Learning," *Microsoft Ignite*, Sep. 2022. https://learn.microsoft.com/en-us/azure/machine-learning/concept-error-analysis (accessed Oct. 06, 2022).

[16] K. N. Vokinger, S. Feuerriegel, and A. S. Kesselheim, "Mitigating bias in machine learning for medicine," *Commun Med*, vol. 1, no. 1, Art. no. 1, Aug. 2021, doi: 10.1038/s43856-021-00028-w.

[17] I. Cofone, "Algorithmic Discrimination Is an Information Problem," *Hastings Law Journal*, vol. 70, no. 6, p. 1389, Aug. 2019.

[18] T. Hellström, V. Dignum, and S. Bensch, "Bias in Machine Learning -- What is it Good for?" arXiv, Sep. 20, 2020. doi: 10.48550/arXiv.2004.00686.

[19] David Danks and Alex John London, "Algorithmic Bias in Autonomous Systems," presented at the International Joint Conference on Artificial Intelligence, 2017. [Online]. Available: https://www.cmu.edu/dietrich/philosophy/docs/london/IJCAI17-AlgorithmicBias-Distrib.pdf

[20] S. Akter, Y. K. Dwivedi, S. Sajib, K. Biswas, R. J. Bandara, and K. Michael, "Algorithmic bias in machine learning-based marketing models," *Journal of Business Research*, vol. 144, pp. 201–216, May 2022, doi: 10.1016/j.jbusres.2022.01.083.

[21] C. N. Heinz, A. Echle, S. Foersch, A. Bychkov, and J. N. Kather, "The future of artificial intelligence in digital pathology – results of a survey across stakeholder groups," *Histopathology*, vol. 80, no. 7, pp. 1121–1127, 2022, doi: 10.1111/his.14659.

[22] T. K. L. Lui *et al.*, "Feedback from artificial intelligence improved the learning of junior endoscopists on histology prediction of gastric lesions," *Endosc Int Open*, vol. 08, no. 2, pp. E139–E146, Feb. 2020, doi: 10.1055/a-1036-6114.

[23] V. Derhami, J. Paksima, and H. Khajeh, "Web pages ranking algorithm based on reinforcement learning and user feedback," *Journal of AI and Data Mining*, vol. 3, no. 2, pp. 157-168., Jun. 2015, doi: 10.5829/idosi.JAIDM.2015.03.02.05.

[24] R. Hafner and M. Riedmiller, "Reinforcement learning in feedback control," *Mach Learn*, vol. 84, no. 1, pp. 137–169, Jul. 2011, doi: 10.1007/s10994-011-5235-x.

[25] J. Huang and C. X. Ling, "Using AUC and accuracy in evaluating learning algorithms," *IEEE Transactions on Knowledge and Data Engineering*, vol. 17, no. 3, pp. 299–310, Mar. 2005, doi: 10.1109/TKDE.2005.50.

[26] S. Liu and L. N. Vicente, "Accuracy and fairness trade-offs in machine learning: a stochastic multi-objective approach," *Comput Manag Sci*, vol. 19, no. 3, pp. 513–537, Jul. 2022, doi: 10.1007/s10287-022-00425-z.

[27] "Accuracy Score," *scikit-learn*, 2022. https://scikit-learn/stable/modules/generated/sklearn.metrics.accuracy_score.html (accessed Oct. 09, 2022).

[28] "Classification: Accuracy | Machine Learning," *Google Developers*. https://developers.google.com/machine-learning/crash-course/classification/accuracy (accessed Oct. 09, 2022).

[29] A. Jobin, M. Ienca, and E. Vayena, "The global landscape of AI ethics guidelines," *Nat Mach Intell*, vol. 1, no. 9, Art. no. 9, Sep. 2019, doi: 10.1038/s42256-019-0088-2.

[30] S. Vollmer *et al.*, "Machine learning and artificial intelligence research for patient benefit: 20 critical questions on transparency, replicability, ethics, and effectiveness," *BMJ*, vol. 368, p. l6927, Mar. 2020, doi: 10.1136/bmj.l6927.

[31] D. S. Char, N. H. Shah, and D. Magnus, "Implementing Machine Learning in Health Care — Addressing Ethical Challenges," *N Engl J Med*, vol. 378, no. 11, pp. 981–983, Mar. 2018, doi: 10.1056/NEJMp1714229.

[32] P. G. John, D. Vijaykeerthy, and D. Saha, "Verifying Individual Fairness in Machine Learning Models," in *Proceedings of the 36th Conference on Uncertainty in Artificial Intelligence (UAI)*, Aug. 2020, pp. 749–758. Accessed: Oct. 09, 2022. [Online]. Available: https://proceedings.mlr.press/v124/george-john20a.html

[33] X. Wang, A. Lu, J. Liu, Z. Kang, and C. Pan, "Intelligent interaction model for battleship control based on the fusion of target intention and operator emotion," *Computers & Electrical Engineering*, vol. 92, p. 107196, Jun. 2021, doi: 10.1016/j.compeleceng.2021.107196.

[34] N. Dimitropoulos, T. Togias, G. Michalos, and S. Makris, "Operator support in human–robot collaborative environments using AI enhanced wearable devices," *Procedia CIRP*, vol. 97, pp. 464–469, Jan. 2021, doi: 10.1016/j.procir.2020.07.006.

[35] Sotiris Kotsiantis, Dimitris Kanellopoulos, and Panayiotis Pintelas, "Handling imbalanced datasets: A review.," *GESTS International Transactions on Computer Science and Engineering,* vol. 30, no. 1, pp. 25–36, 2006.

[36] G. Haixiang, L. Yijing, J. Shang, G. Mingyun, H. Yuanyue, and G. Bing, "Learning from class-imbalanced data: Review of methods and applications," *Expert Systems with Applications*, vol. 73, pp. 220–239, May 2017, doi: 10.1016/j.eswa.2016.12.035.

[37] H. He and E. A. Garcia, "Learning from Imbalanced Data," *IEEE Transactions on Knowledge and Data Engineering*, vol. 21, no. 9, pp. 1263–1284, Sep. 2009, doi: 10.1109/TKDE.2008.239.

[38] J. G. Moreno-Torres, T. Raeder, R. Alaiz-Rodríguez, N. V. Chawla, and F. Herrera, "A unifying view on dataset shift in classification," *Pattern Recognition*, vol. 45, no. 1, pp. 521–530, Jan. 2012, doi: 10.1016/j.patcog.2011.06.019.

[39] R. V. Yampolskiy, "Unpredictability of AI." arXiv, May 29, 2019. doi: 10.48550/arXiv.1905.13053.

[40] United Nations Institute for Disarmament Research and A. Holland Michel, "The Black Box, Unlocked: Predictability and Understandability in Military AI." United Nations Institute for Disarmament Research, Sep. 2020. [Online]. Available: 10.37559/SecTec/20/AI1

[41] S. Ochella and M. Shafiee, "Performance Metrics for Artificial Intelligence (AI) Algorithms Adopted in Prognostics and Health Management (PHM) of Mechanical Systems," *J. Phys.: Conf. Ser.*, vol. 1828, no. 1, p. 012005, Feb. 2021, doi: 10.1088/1742-6596/1828/1/012005.

[42] W. Lehnert and B. Sundheim, "A Performance Evaluation of Text-Analysis Technologies," *AI Magazine*, vol. 12, no. 3, Art. no. 3, Sep. 1991, doi: 10.1609/aimag.v12i3.905.

[43] K. Yu *et al.*, "Design and Performance Evaluation of an AI-Based W-Band Suspicious Object Detection System for Moving Persons in the IoT Paradigm," *IEEE Access*, vol. 8, pp. 81378–81393, 2020, doi: 10.1109/ACCESS.2020.2991225.

[44] H. Xu and S. Mannor, "Robustness and generalization," *Mach Learn*, vol. 86, no. 3, pp. 391–423, Mar. 2012, doi: 10.1007/s10994-011-5268-1.

[45] "Machine Learning Robustness: New Challenges and Approaches - Vector Institute for Artificial Intelligence," Mar. 29, 2022. https://vectorinstitute.ai/2022/03/29/machine-learning-robustness-new-challenges-and-approaches/ (accessed Oct. 09, 2022).

[46] A. F. Cooper, E. Moss, B. Laufer, and H. Nissenbaum, "Accountability in an Algorithmic Society: Relationality, Responsibility, and Robustness in Machine Learning," in *2022 ACM Conference on Fairness, Accountability, and Transparency*, New York, NY, USA, Jun. 2022, pp. 864–876. doi: 10.1145/3531146.3533150.

[47] L. Hancox-Li, "Robustness in machine learning explanations: does it matter?," in *Proceedings of the 2020 Conference on Fairness, Accountability, and Transparency*, New York, NY, USA, Jan. 2020, pp. 640–647. doi: 10.1145/3351095.3372836.

[48] D. Didona, F. Quaglia, P. Romano, and E. Torre, "Enhancing Performance Prediction Robustness by Combining Analytical Modeling and Machine Learning," in *Proceedings of the 6th ACM/SPEC International Conference on Performance Engineering*, New York, NY, USA, Jan. 2015, pp. 145–156. doi: 10.1145/2668930.2688047.

[49] R. Mangal, A. V. Nori, and A. Orso, "Robustness of Neural Networks: A Probabilistic and Practical Approach." arXiv, Feb. 15, 2019. doi: 10.48550/arXiv.1902.05983.

[50] B. Biggio, B. Nelson, and P. Laskov, "Poisoning Attacks against Support Vector Machines." arXiv, Mar. 25, 2013. doi: 10.48550/arXiv.1206.6389.

[51] A. N. D. Meyer, T. D. Giardina, C. Spitzmueller, U. Shahid, T. M. T. Scott, and H. Singh, "Patient Perspectives on the Usefulness of an Artificial Intelligence-Assisted Symptom Checker: Cross-Sectional Survey Study," *J Med Internet Res*, vol. 22, no. 1, p. e14679, Jan. 2020, doi: 10.2196/14679.

[52] J. H. Kim, S.-J. Nam, and S. C. Park, "Usefulness of artificial intelligence in gastric neoplasms," *World Journal of Gastroenterology*, vol. 27, no. 24, pp. 3543–3555, Jun. 2021, doi: 10.3748/wjg.v27.i24.3543.

[53] T. B. of C. Secretariat, "Directive on Automated Decision-Making," Feb. 05, 2019. https://www.tbs-sct.canada.ca/pol/doc-eng.aspx?id=32592#appA (accessed Oct. 09, 2022).

[54] UNESCO, "UNESCO SCIENCE REPORT The race against time for smarter development," United Nations Educational, Scientific and Cultural Organization, France, 2021. Accessed: Oct. 09, 2022. [Online]. Available: https://unesdoc.unesco.org/ark:/48223/pf0000377433/PDF/377433eng.pdf.multi

[55] MINISTER OF INNOVATION, SCIENCE AND INDUSTRY, "Government Bill (House of Commons) C-27 (44-1) - First Reading - Digital Charter Implementation Act, 2022 - Parliament of Canada." https://www.parl.ca/DocumentViewer/en/44-1/bill/C-27/first-reading (accessed Oct. 09, 2022).

[56] A. F. T. Winfield *et al.*, "IEEE P7001: A Proposed Standard on Transparency," *Frontiers in Robotics and AI*, vol. 8, 2021, Accessed: Oct. 09, 2022. [Online]. Available: https://www.frontiersin.org/articles/10.3389/frobt.2021.665729

[57] U. Bhatt *et al.*, "Explainable machine learning in deployment," in *Proceedings of the 2020 Conference on Fairness, Accountability, and Transparency*, New York, NY, USA, Jan. 2020, pp. 648–657. doi: 10.1145/3351095.3375624.

[58] S. Larsson and F. Heintz, "Transparency in artificial intelligence," *Internet Policy Review*, vol. 9, no. 2, May 2020, Accessed: Oct. 09, 2022. [Online]. Available: https://policyreview.info/concepts/transparency-artificial-intelligence

[59] Y. Hu *et al.*, "Artificial Intelligence Security: Threats and Countermeasures," *ACM Comput. Surv.*, vol. 55, no. 1, p. 20:1-20:36, Nov. 2021, doi: 10.1145/3487890.

[60] A. Ziller, J. Passerat-Palmbach, A. Trask, R. Braren, D. Rueckert, and G. Kaissis, "Artificial Intelligence in Medicine and Privacy Preservation," in *Artificial Intelligence in Medicine*, N. Lidströmer and H. Ashrafian, Eds. Cham: Springer International Publishing, 2020, pp. 1–14. doi: 10.1007/978-3-030-58080-3_261-1.

[61] W. N. Price and I. G. Cohen, "Privacy in the age of medical big data," *Nat Med*, vol. 25, no. 1, pp. 37–43, Jan. 2019, doi: 10.1038/s41591-018-0272-7.

[62] M. Ryan, "In AI We Trust: Ethics, Artificial Intelligence, and Reliability," *Sci Eng Ethics*, vol. 26, no. 5, pp. 2749–2767, Oct. 2020, doi: 10.1007/s11948-020-00228-y.

[63] A. Oseni, N. Moustafa, H. Janicke, P. Liu, Z. Tari, and A. Vasilakos, "Security and Privacy for Artificial Intelligence: Opportunities and Challenges." arXiv, Feb. 09, 2021. doi: 10.48550/arXiv.2102.04661.

[64] M. Krotofil, J. Larsen, and D. Gollmann, "The Process Matters: Ensuring Data Veracity in Cyber-Physical Systems," in *Proceedings of the 10th ACM Symposium on Information, Computer and Communications Security*, New York, NY, USA, Apr. 2015, pp. 133–144. doi: 10.1145/2714576.2714599.

[65] T. Lukoianove and V. Rubin, "Veracity Roadmap: Is Big Data Objective, Truthful and Credible?," *ASIST 2013: 76th Annual Meeting of the Association for Information Science and Technology: Beyond the Cloud: Rethinking Information Boundaries*, vol. 24, Nov. 2013, doi: 10.7152/acro.v24i1.14671.

[66] "IEEE P2802 - Standard for the Performance and Safety Evaluation of Artificial Intelligence Based Medical Device: Terminology," 2022. https://www.standict.eu/standards-repository/ieee-p2802-standard-performance-and-safety-evaluation-artificial-intelligence (accessed Oct. 12, 2022).

[67] Calvin Zhu and Thomas E Doyle, "Ultrasound Operator Variance Classification for Agency in Artificial Intelligence Support of Cyber-Physical Systems," presented at the 2022 IEEE Canadian Conference on Electrical and Computer Engineering (CCECE), Sep. 2022.

[68] A. Simons, T. Doyle, D. Musson, and J. Reilly, "Impact of Physiological Sensor Variance on Machine Learning Algorithms," in *2020 IEEE International Conference on Systems, Man, and Cybernetics (SMC)*, Oct. 2020, pp. 241–247. doi: 10.1109/SMC42975.2020.9282912.

[69] T. Gehr, M. Mirman, D. Drachsler-Cohen, P. Tsankov, S. Chaudhuri, and M. Vechev, "AI2: Safety and Robustness Certification of Neural Networks with Abstract Interpretation," in *2018 IEEE Symposium on Security and Privacy (SP)*, May 2018, pp. 3–18. doi: 10.1109/SP.2018.00058.